\documentclass[conference]{IEEEtran}
\IEEEoverridecommandlockouts
\usepackage{cite}
\usepackage{amsmath,amssymb,amsfonts}
\usepackage{algorithmic}
\usepackage{graphicx}
\usepackage{textcomp}
\usepackage{xcolor}
\usepackage{svg}
\usepackage{amsthm}
\usepackage{comment}
\usepackage{booktabs}
\usepackage{colortbl} % to color table cells
\usepackage{url}

\newtheorem{definition}{Definition}
\newtheorem{defn}{Definition}[]

\newcommand{\sectionname}{Sec.}
\def\definitionname{Definition}
\newcommand{\colorcell}{ \cellcolor{gray!35} }
\mathchardef\mhyphen="2D % Short hypen

\def\BibTeX{{\rm B\kern-.05em{\sc i\kern-.025em b}\kern-.08em
    T\kern-.1667em\lower.7ex\hbox{E}\kern-.125emX}}

\begin{document}

\title{What's Coming Next? Short-Term Simulation of\\Business Processes from Current State\\
}

\author{\IEEEauthorblockN{1\textsuperscript{st} Maksym Avramenko}
\IEEEauthorblockA{\textit{University of Tartu} \\
maksym.avramenko@ut.ee}
\and
\IEEEauthorblockN{2\textsuperscript{nd} David Chapela-Campa}
\IEEEauthorblockA{\textit{University of Tartu} \\
david.chapela@ut.ee}
\and
\IEEEauthorblockN{3\textsuperscript{th} Marlon Dumas}
\IEEEauthorblockA{\textit{University of Tartu} \\
marlon.dumas@ut.ee}
\and
\IEEEauthorblockN{4\textsuperscript{rd} Fredrik Milani}
\IEEEauthorblockA{\textit{University of Tartu} \\
fredrik.milani@ut.ee}
% \and
% \and
}

\maketitle

\begin{abstract}
Business process simulation is an approach to evaluate business process changes prior to implementation.
Existing methods in this field primarily support tactical decision-making, where simulations start from an empty state and aim to estimate the long-term effects of process changes. 
A complementary use-case is operational decision-making, where the goal is to forecast short-term performance based on ongoing cases and to analyze the impact of temporary disruptions, such as demand spikes and shortfalls in available resources.
An approach to tackle this use-case is to run a long-term simulation up to a point where the workload is similar to the current one (warm-up), and measure performance thereon. 
However, this approach does not consider the current state of ongoing cases and resources in the process. 
This paper studies an alternative approach that initializes the simulation from a representation of the current state derived from an event log of ongoing cases.
The paper addresses two challenges in operationalizing this approach: (1) Given a simulation model, what information is needed so that a simulation run can start from the current state of cases and resources? (2) How can the current state of a process be derived from an event log?
The resulting short-term simulation approach is embodied in a simulation engine that takes as input a simulation model and a log of ongoing cases, and simulates cases for a given time horizon. An experimental evaluation shows that this approach yields more accurate short-term performance forecasts than long-term simulations with warm-up period, particularly in the presence of concept drift or bursty performance patterns.
\end{abstract}


%!TEX root = main.tex

\section{Introduction\label{sec:introduction}}

% Introduction of the topic (generally define BPS and why is useful) %,DBLP:books/sp/DumasRMR18
Business Process Simulation (BPS) enables analysts to evaluate how business process changes affect one or more performance metrics such as resource utilization and cycle time~\cite{DBLP:series/ihis/Aalst15}.
%It allows organizations to explore ``what-if'' scenarios by estimating the effects of process changes prior to their implementation.
%This simulation-based approach can lead to significant cost savings and efficiency gains by enabling data-driven decision-making~\cite{DBLP:books/sp/DumasRMR18}.
% Introduce short-term purpose and what it is about.
Most BPS approaches target tactical decision-making use cases. They enable analysts to estimate the impact of changes on the overall performance of the process, typically simulating over a time horizon of weeks or months. For example, an analyst might use BPS to estimate how much cycle time can be reduced by automating a verification task. % (much longer than the cycle time of the process)
%~\cite{DBLP:books/sp/DumasRMR18}.

%estimate how would a long-term simulation, where the estimations are based on average performance metrics over the simulated period.
A complementary use case arises in operational decision-making~\cite{ReijersAalst1999}, where managers want to assess how changes or disruptions, such as spikes in the number of cases or staff reallocation, affect both ongoing cases and those that will begin within a time frame similar to the process’s cycle time.
%In contrast, \emph{short-term simulation} aims to optimize processes in an operational setting by using up-to-date data from ongoing operations
%, focusing on near-term time periods such as a few weeks. % or a few months.
%This approach is particularly useful for operational decision-making, where quick adjustments can lead to immediate improvements in performance, such as reallocating resources to address an unexpected surge in demand~\cite{DBLP:journals/dke/RozinatWAHF09,DBLP:conf/bpm/WynnDFHA07}.% Introduction of the problem
Simulations over short time periods (short-term simulations) start from a state identical or similar to the current one~\cite{ReijersAalst1999}. In this setting, the current state of an ongoing process refers to, for instance, the control-flow state of each ongoing case (i.e.,\ which activities have been, are being, and are ready to be executed), the resource allocation, and data attribute values.
%Otherwise, short-term simulations lead to inaccurate predictions~\cite{DBLP:journals/dke/RozinatWAHF09,DBLP:conf/bpm/WynnDFHA07}.
%may not reflect the actual conditions of the process, which can
%In the context of short-term simulation, accurately capturing the \emph{current state} of ongoing cases becomes crucial.

% % Previous work

An approach to short-term simulation is to run a long-term simulation starting from an empty state (no cases), up to the point where the workload stops increasing (\emph{warm-up} method). The state where the workload stops increasing is referred to as a \emph{steady state})~\cite{DBLP:series/ihis/Aalst15}.\footnote{Measuring the workload, for example, as the number of active cases.}
From this steady state on, we can measure the performance of the process for the desired time horizon (e.g., one week).
%common method to avoid the impact of starting the simulation in an empty state is to discard an initial portion of the simulation  to allow the process to reach a.
%A steady state typically refers to a stable operating condition in which arrival rates, resource utilization, and other performance measures vary around an equilibrium, rather than changing dramatically over time. 
%Once this period ends, any initial transient behavior is discarded, and performance metrics are collected on the assumed ``stable'' portion of the simulation.
This approach presupposes that once a steady state is reached, the performance does not vary over time. This assumption does not hold when the workload fluctuates due to, for example, periodic spikes (e.g.,\ at the start of each month) or unforeseen spikes (e.g.,\ marketing campaigns)~\cite{DBLP:conf/bpm/PourbafraniLLA23}. In such scenarios, the reached steady state after the warm-up period may differ from the current state. A variation of this approach is to run the simulation until the workload is similar to the current one. However, such a state might not be reached, e.g.,\ if the current state has a higher workload than the simulation model can replicate (concept drift).

An alternative approach is to initialize the simulation state so that it mirrors the current process state. This method, as proposed in~\cite{DBLP:journals/dke/RozinatWAHF09,DBLP:conf/bpm/WynnDFHA07}, applies when a workflow management system records the current state of each enabled or ongoing activity instance, as well as the status of each resource (whether busy or available). In this paper, we tackle the problem of initializing a simulation to mirror a current state in the more general setting where a process is executed across one or more systems that do not store information about enabled activity instances and resources. Instead, we take a BPS model and an event log of ongoing cases as a starting point.
%practice, many processes are not executed on such systems, but rather occur by virtue of users performing work on one or more enterprise systems, which do not explicitly keep track of all activity instances.
We propose a method that, from the event log, derives a representation of the state of the process w.r.t.\ the given BPS model. We, then, load this state into a simulation engine and run simulations starting from the current state for a specified time horizon.
%First, we analyze the information needed for a simulation to start from the current state. Accordingly, we formalize the components of the state of a business process simulation. Second, we propose a method to extract such a state directly from an event log containing ongoing cases.
%Rather than relying on a warm-up period or assumptions of past executions, we identify the key components of the state -- i.e., ongoing activities, enabled activities, and current resource allocations -- and feed them to the simulation engine.
%This allows more accurate reflection of \emph{current} conditions, ultimately improving short-term decision-making.
%Our hypothesis is that starting the simulation in the exact current state of the process leads to more accurate predictions of the near-future process execution.
The paper reports on an experimental evaluation that compares the accuracy of the proposed approach relative to a warm-up approach, in the context of processes with a stable steady state, a bursty steady state, as well as processes with concept drift.

%The evaluation relies on 4 event logs of 4 synthetic processes of different complexities and workload dynamics, and 4 real-life event logs with and without concept drift.

%evaluate our approach, we conduct experiments using both synthetic and real-world event logs. 

% The evaluation demonstrates the effectiveness of our method in capturing the current state and producing realistic simulation outcomes.

% % Outline 
% The paper is structured as follows...

%The remainder of the paper is structured as follows:
%Section~\ref{sec:background} provides an overview of relevant concepts and related work.
%Section~\ref{sec:approach} details our proposed method for extracting the process state.
%Section~\ref{sec:evaluation} presents the evaluation results, and Section~\ref{sec:conclusion} concludes the paper and discusses future research directions.

%!TEX root = main.tex

%\section{Background and Related Work}

\section{Related Work\label{sec:related-work}}

% Short paragraph about ReijersAalst1999 just to present that the concept was firstly introduced by Reijers and Aalst in 99, they talk about it, blabla.
The application of BPS for operational decision-making was introduced in~\cite{ReijersAalst1999}.
The authors proposed using data from enterprise systems, such as Workflow Management (WFMS) or Enterprise Resource Planning (ERP) systems, to discover BPS models that reflect the execution of the process and extract the state of the ongoing process at the starting point of the simulation.
Building on this idea, Wynn et al.~\cite{DBLP:conf/bpm/WynnDFHA07} and Rozinat et al.~\cite{DBLP:journals/dke/RozinatWAHF09} propose a system architecture to discover a BPS model from traces of completed cases and to extract the state of ongoing cases. 
Liu et al.~\cite{DBLP:journals/dss/LiuZLJ12} propose a similar architecture that builds a BPS model based on event graphs instead of Petri nets. 
These studies assume that the process is executed in a WFMS that explicitly records the current state of each ongoing case (e.g., which activity instances are enabled and when they were enabled, and which resources are available and since when). 
In this paper, we target the scenario where the available information consists of trace prefixes of ongoing cases stored in an event log (i.e., there is no WFMS available), from which the current state needs to be derived.

Pourbafrani et al.~\cite{DBLP:conf/bpm/PourbafraniLLA23} introduced a data-driven method to estimate a \emph{steady state} from historical event logs to initialize simulations (both long-term and short-term). Their approach takes as input a periodicity factor (e.g., weekly) specified by an expert. The approach uses a log of completed cases to estimate an ``average process state'' for the given periodicity factor, e.g.\ the average process state on Mondays at 8AM. The assumption is that the workload of the process is periodic, e.g. the workload next Monday at 8am is similar to the workload on previous Mondays at 8am. The approach we propose in this paper does not make this assumption, and hence, it can handle irregular workloads and concept drifts (e.g.\ workloads not previously observed). Importantly, the approach in~\cite{DBLP:conf/bpm/PourbafraniLLA23} only estimates the control-flow state, that is, it computes a state consisting of a set of activity instances that may be ongoing. However, it does not calculate the \emph{enablement times} of these activity instances. Thus, the output of~\cite{DBLP:conf/bpm/PourbafraniLLA23} cannot be used to initialize the activity instance queues for short-term simulations. Furthermore, their approach does not take into account  resources, i.e.\ it does not determine which resources are busy and which ones are available in the current state. The approach we propose in this paper addresses these limitations.

%To address these limitations, we propose an alternative approach that takes a BPS model (representing the as-is process or a what-if scenario with changes) and a log of ongoing cases as input. This approach not only identifies the ongoing activities but also determines which activities are currently enabled, along with their \emph{enablement times}, enabling accurate initialization of activity queues for simulations.

Kraus et al.\cite{caise-kraus-steady-state-ppm} highlighted the importance of steady states in predictive process monitoring (PPM), proposing a method to identify historical log periods where the process workload remains steady. Unlike approaches that track the state at a point in time, this focuses on finding these stable periods. However, while PPM methods\cite{DBLP:conf/caise/MaggiFDG14,DBLP:journals/is/SenderovichFM19} can predict the performance of current and upcoming cases in an as-is scenario, they are not suitable for assessing the impact of changes to the process (i.e., a to-be configuration).
%~\cite{DBLP:conf/caise/MaggiFDG14,DBLP:conf/bpm/Francescomarino18,DBLP:journals/tsc/Marquez-Chamorro18,DBLP:journals/tkdd/TeinemaaDRM19,DBLP:journals/is/SenderovichFM19}

%Short-term simulation has been applied within the healthcare industry. For instance, Pegoraro et al.,~\cite{DBLP:conf/worldcist/PegoraroSLDSC18} propose applying short-term BPS for healthcare management processes. Bahrani et al.~\cite{DBLP:conf/scsc/BahraniTMA13} use short-term BPS for resource allocation to assess impact on wait times or to analyze resource allocation with the objective of reducing patient length-of-stay~\cite{DBLP:journals/simulation/Bedoya-Valencia16}.
%Although these works showcase the utility of short-term simulation in different domains, they do not address discovering the state of an ongoing process.

Finally, the application of short-term simulation for operational decision-making has also been studied within various domains.
For instance, Bahrani et al.~\cite{DBLP:conf/scsc/BahraniTMA13} and Bedoya-Valencia et al.~\cite{DBLP:journals/simulation/Bedoya-Valencia16} studied the use of short-term simulation to predict the impact on waiting times (e.g., patient length-of-stay) of different resource allocation policies in healthcare processes.
% Bahrani et al.~\cite{DBLP:conf/scsc/BahraniTMA13} the use of short-term simulation to predict the impact on the waiting times of different resource allocations in the Canadian healthcare system.
% Bedoya-Valencia et al.~\cite{DBLP:journals/simulation/Bedoya-Valencia16} propose to use short-term simulation to analyze resource allocation with the objective of reducing patient length-of-stay (LOS) while leveling resource utilization.
Similarly, Pegoraro et al.\ proposed in~\cite{DBLP:conf/worldcist/PegoraroSLDSC18} to apply short-term BPS to improve healthcare management processes.
Khodyrev et al.~\cite{DBLP:conf/iccS/KhodyrevP14} studied the application of BPS for KPI prediction in a short-term setting of industrial repair processes.
Although these works showcase the utility of short-term simulation in different domains, they do not address the problem of discovering the current state of an ongoing process from data.

% Our proposal focuses on...
%In contrast, our work focuses on directly capturing the current state of the process from an event log of ongoing cases.
%This allows for a more accurate initialization of simulations, robust to recent concept drifts, better reflecting real-time conditions, and enhances the relevance of the simulation results for operational decision-making.

\section{Preliminaries\label{sec:background}}

%\todo[inline]{[come up with] notion of a case that is not deterministically finished. If the last activity may occur also in the middle of the trace, we might assume the case is not finished. Define simulation start, simulation horizon, simulation period.}

%A \emph{business process} is a set of interrelated activities $\mathcal{A}$ designed to achieve a specific organizational goal, such as processing an order, approving a loan, or handling a customer complaint~\cite{DBLP:books/sp/DumasRMR18}.
%These activities are typically structured and may involve decision points, task sequences, and resource allocations.
As a starting point, we take a business process whose execution is recorded in an \emph{event log}, consisting of a set of events, each of which records facts about an activity instance~\cite{DBLP:journals/tkde/AalstWM04}. 
%We consider a type of event log where each event captures all available information about an activity instance (\emph{activity instance log}).
Let $\alpha \in \mathcal{A}$ denote the label of an \emph{activity} of the process.
An \emph{activity instance} $\varepsilon = (\phi, \alpha, \tau_{s}, \tau_{c}, \rho)$ denotes the execution of the activity $\alpha$, where 
$\phi$ represents the \emph{case identifier},
%$\tau_{e}$ is the \emph{enablement time}, indicating when the activity is ready to start.
$\tau_{s}$ and $\tau_{c}$ represent the \emph{start time} and \emph{completion time} of the activity instance,
and $\rho$ denotes the \emph{resource} assigned to perform the activity.
Accordingly, we write $\phi(\varepsilon_{i})$, $\alpha(\varepsilon_{i})$, $\tau_{s}(\varepsilon_{i})$, $\tau_{c}(\varepsilon_{i})$, and $\rho(\varepsilon_{i})$ to denote, respectively, the case, the activity, the start and end timestamps, and the resource associated with the activity instance $\varepsilon_{i}$. 
The end time of an activity instance may store a null value  (``-'') to denote that its end time is unknown as of the largest time recorded in the log.
Finally, although not included in the activity instance definition, we use $\tau_{e}(\varepsilon)$ to denote the time instant when an activity became \emph{enabled}, i.e., available for processing.

A \emph{trace} is a sequence of events $\Upsilon = \langle \varepsilon_1, \varepsilon_2, \ldots, \varepsilon_m \rangle$, where $\forall_{i,j \in 1..m} : \phi(\varepsilon_i) = \phi(\varepsilon_j)$, and such that the events are sorted in ascending order $\tau_{s}(\varepsilon_i) \leq \tau_{s}(\varepsilon_j)$ for $1 \leq i < j \leq m$.
%(or $\tau_c$ if the start times are missing). as a sequence of events corresponding to the same process instance (or \emph{case}).
%A trace is thus a sequence of events 
%\[
%\Upsilon = \langle \varepsilon_1, \varepsilon_2, \ldots, \varepsilon_m \rangle 
%\]
An \emph{event log} $\mathcal{L} = \{\Upsilon_1, \Upsilon_2, \dots, \Upsilon_k\}$ is a collection of traces, where
%\[
%\mathcal{L} = \{\Upsilon_1, \Upsilon_2, \dots, \Upsilon_k\}.
%\]
each trace $\Upsilon_i$ corresponds to a \emph{case} (i.e., a \emph{process instance}), capturing all the activity instances executed from the first until the last recorded event in that case.
An \emph{ongoing case} is a case containing ongoing activity instances, or such that some of its events are not recorded, as of the largest timestamp in a log.
The trace of an ongoing case is called its \emph{trace prefix}.
An \emph{ongoing cases log} is an event log containing only trace prefixes.\footnote{
  We do not address the question of determining if a case is ongoing given a trace thereof. Instead, we assume that every trace in the input log is ongoing.
}
  %There is no single way to automatically assess whether a process case is terminated or not without the presence of a final activity in the process model~\cite{DBLP:journals/tsc/chapela2025efficient}.
%A \emph{case} is said to be \emph{ongoing} if it has not reached any known completion condition.
%For example, if the process model indicates that ``Notify Customer'' is the final task, but the log for this case does not contain an event for that final task (or it contains it partially without an end time), we treat the case as ongoing. 
%In other words, an ongoing case is one whose recorded trace does not yet match a complete path from start to end event in the BPMN model.
Table~\ref{tab:event_log_interleaved} depicts an ongoing cases log. 
%Each row represents an activity instance that records the execution of a specific activity within a process instance (or case). This information is crucial for analyzing process performance, identifying bottlenecks, and discovering patterns in process executions.

% In this log, each case is incomplete:
% \begin{itemize}
%     \item \textbf{Case 101}: Stuck in the parallel branch, with ``Prepare Package'' started but not finished.
%     \item \textbf{Case 102}: Also in the parallel branch, but the packaging and invoice tasks are largely done—except ``Send Invoice'' has no end time.
%     \item \textbf{Case 103}: All tasks have finished except ``Ship Order,'' which has not yet started (this final activity is pending).
% \end{itemize} 
% In later sections, we will use such partially completed cases to illustrate how our method discovers the \emph{ongoing state} and resumes the simulation from that point.

A Business Process Simulation model (BPS model) consists of a \emph{process model} and a set of simulation parameters (discussed below).
%is a graphical representation of the set of traces of a process commonly used to understand, analyze, and optimize how tasks are executed within an organization~\cite{DBLP:journals/is/RozinatA08}.
%There are many formalisms to represent process models, e.g., BPMN\footnote{\url{https://www.bpmn.org/}} or Petri nets~\cite{DBLP:conf/bpm/Aalst00}.
We represent process models as \emph{workflow graphs}, which capture a subset of BPMN~\cite{DBLP:journals/is/FavreFV15}. 
%We define a workflow graph as follows:

\begin{definition}[Workflow graph, based on~\cite{DBLP:journals/is/FavreFV15}\label{def:workflow-graph}]
    A Workflow graph (WF-graph) is a directed graph $W = (start, sink, N, F, t)$, where:
    \begin{itemize}
        \item $start$ denotes the \textbf{start event} of the process;
        \item $sink$ denotes the \textbf{end event} of the process;
        \item $N$ is a set of \textbf{nodes} $n$, representing either tasks (activities), events, or gateways;
        \item $F \subseteq (N \cup \{ start \}) \times (N \cup \{sink \})$ is a set of \textbf{flows} $f$, denoting the transitions between nodes; and
        \item $t: N \rightarrow \{\text{task}, \text{event}, \text{AND}, \text{XOR}\}$ is a function that assigns a \textbf{type} to each node, indicating whether it is an activity, and event, or a gateway.\footnote{Unlike~\cite{DBLP:journals/is/FavreFV15}, we consider only workflow graphs without inclusive (IOR) gateways. An extension to support IOR gateways is left as future work.}
    \end{itemize}
%    $g: F \rightarrow (N \cup \{start\}) \times (N \cup \{end\})$ is a function that assigns each flow to a pair of nodes, specifying the source and target of the flow.
    %The sets $n\bullet$ and $\bullet n$ denote, respectively, the outgoing and incoming flows for a node $n \in N$.
    %The sets $start\bullet$ and $\bullet sink$ refer, respectively, to the outgoing flow of the start event and the incoming flow of the end event.

    For a given flow $f \in F$, let $source(f)$ be the source of this flow. Given a WF-Graph $W = (start, sink, N, F, t)$, we define its set of activities as $A(W) = \{ a \in N \mid t(a) = \mathit{task} \}$, its set of intermediate events as $E(W) = \{ e \in N \mid t(e) = \mathit{event} \}$, and its conditional flows as $CF(W) = \{ f \in F \mid x = source(f) \wedge t(x) = \mathit{XOR} \wedge ( \exists f' \in F: f' \neq f \wedge x = source(f') ) \}$, in other words, the source of a conditional flow is an XOR gateway that has at least one other outgoing flow. %Where there is no ambiguity, we will simply write $A$, $E$, and $\mathit{CF}$ to refer to the activities, events, and conditional flows of a workflow graph.
\end{definition}

\begin{table}[t]
    \vspace*{-3mm}
    \centering
    \scriptsize
    \caption{Example of an ongoing cases log.}
    \label{tab:event_log_interleaved}
    \begin{tabular}{l l l l l}
    \toprule
    \textbf{Case ID} & \textbf{Activity} & \textbf{Start Time} & \textbf{End Time} & \textbf{Resource} \\
    \midrule

    101 & Collect Customer Info.      & 08:00 & 08:10 & Alice \\
    101 & Collect Express Payment     & 08:15 & 08:25 & Alice \\
    101 & Prepare Package             & 08:30 & -     & Bob \\

    \midrule
    102 & Collect Customer Info.      & 08:20 & 08:35 & Dave \\
    102 & Collect Standard Payment    & 08:40 & 08:55 & Dave \\
    102 & Prepare Package             & 09:00 & 09:20 & Bob \\
    102 & Prepare Invoice             & 09:00 & 09:10 & Carol \\
    102 & Send Invoice                & 09:10 & -     & Carol \\

    \midrule
    103 & Collect Customer Info.      & 09:05 & 09:15 & Alice \\
    103 & Collect Express Payment     & 09:20 & 09:30 & Alice \\
    103 & Prepare Package             & 09:35 & 09:50 & Bob \\
    103 & Prepare Invoice             & 09:35 & 09:45 & Carol \\
    103 & Send Invoice                & 09:45 & 09:55 & Carol \\

    \bottomrule
    \end{tabular}
    \vspace*{-3mm}
\end{table}

Similar to Petri nets, the semantics of workflow graphs are defined via a token game where tokens move when nodes are executed. The execution of a case starts with a single token in the start event and ends with a single token in the end event.\footnote{We focus on \emph{sound} workflow graphs as defined in~\cite{DBLP:journals/is/FavreFV15}.} 
%, which  follow the same soundness properties as described for workflow nets in~\cite{DBLP:conf/bpm/Aalst00}
A \emph{marking} $M = \{ f \mid f \in F\}$ is a set of active flows (i.e., having a token) that represents a state in the control-flow execution of the process.
Given a marking $M$, a node of type ``AND'' is \emph{enabled} to be executed iff $M$ contains all its incoming flows, and its execution removes all these flows from the set $M$ and adds all its outgoing flows to $M$.
A node of type ``XOR'' is \emph{enabled} iff $M$ contains one of its incoming flows, and its execution removes this flow from $M$ and adds one of its outgoing flows to $M$.
Finally, the execution of a node of type ``task'' or ``event'' is \emph{enabled} iff $M$ contains at least one of its incoming flows and its execution removes this flow from $M$ and adds its outgoing flows to $M$.
%, or ``XOR'' adds one of its outgoing flows to $M$; and the execution of a node of type ``AND'' adds all its outgoing flows to $M$.
%Given a marking $M_{1}$, a node $n \in N$ of type ``task'', ``event'', or ``AND'' is \emph{enabled} to be executed iff $\bullet n \subseteq M_{1}$, and its execution produces the marking $M_{2} = M_{1} \setminus \bullet n \cup n \bullet$.
%In the same context, a node $n \in N$ of type ``XOR'' is enabled iff $M_{1} \cap \bullet n \neq \emptyset$, and its execution produces the marking $M_{2} = M_{1} \setminus \bullet n \cup \{f_{1}\}$ such that $f_{1} \in n \bullet$.
%Following the notation in~\cite{DBLP:conf/bpm/Aalst00}, given a marking $M_{1}$, we use $M_{1} \xrightarrow{n} M_{2}$ to denote that $n$ is enabled in $M_{1}$, and its execution results in $M_{2}$.
%We use $M_{1} \xrightarrow{\sigma} M_{2}$ to denote that executing the sequence of nodes $\sigma = n_{1},n_{2},\ldots,n_{m}$ from marking $M_{1}$ leads to marking $M_{2}$.
%Finally, a marking $M_{2}$ is said to be a \emph{reachable marking} from $M_{1}$ iff there exists a sequence (possibly empty) $\sigma$ of nodes such that $M_{1} \xrightarrow{\sigma} M_{2}$.

\begin{figure}[t]
\vspace*{-3mm}
    \centering
    \includegraphics[width=0.95\columnwidth]{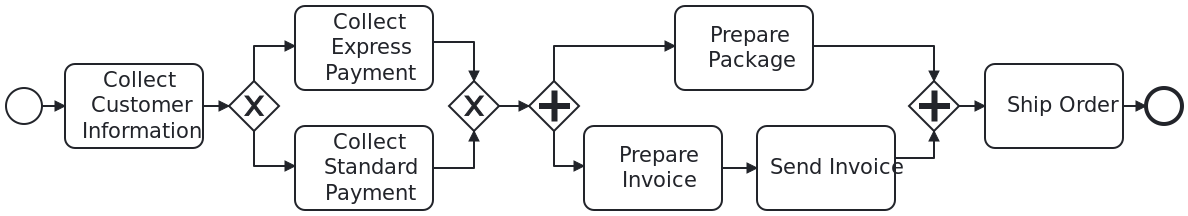}
    \caption{WF-graph of an Order Handling process.}
    \label{fig.bpmn-model}
\vspace*{-3mm}
\end{figure}

Figure~\ref{fig.bpmn-model} depicts a WF-graph using BPMN graphical conventions for events, tasks, and gateways. %for an \emph{order handling process} 

%that includes both an exclusive (XOR) gateway and a parallel (AND) gateway. The model consists of several components:
%\begin{itemize}
%    \item \textbf{Activities}: 
%    \begin{itemize}
%        \item \emph{Collect Customer Information} (initial data gathering),
%        \item \emph{Collect Express Payment} / \emph{Collect Standard Payment} (mutually exclusive payment modes),
%        \item \emph{Prepare Package} (packaging tasks),
%        \item \emph{Prepare Invoice} and \emph{Send Invoice} (invoice-related tasks done in parallel with packaging),
%        \item \emph{Ship Order} (final activity).
%    \end{itemize}
%    \item \textbf{Gateways}:
%    \begin{itemize}
%        \item An \textbf{XOR gateway} decides whether the order requires express or standard payment.
%        \item An \textbf{AND gateway} splits the process into two parallel branches (``Prepare Package'' vs.\ ``Prepare/Send Invoice'') and later joins them before shipping.
%    \end{itemize}
%    \item \textbf{Transitions}: Arrows represent the sequence flow between tasks, gateways, and events.
%    \item \textbf{Events}: The process has one \emph{Start Event} and one \emph{End Event}.
%\end{itemize}

%This BPMN model defines the structure of the process, allowing for a clear understanding of the sequence of activities and decision points involved in handling orders. However, to analyze process performance or predict outcomes, it is essential to extend the process model with additional information for simulation~\cite{DBLP:series/ihis/Aalst15}.

A \emph{Business Process Simulation Model} (BPSM) extends a WF-graph with parameters that allow for the replication of the execution of activity instances and waiting times between these executions. 
%Below, we formalize the components of a BPSM.

\begin{definition}[Business Process Simulation Model\label{def:bps-model}]
A Business Process Simulation Model (BPSM) is a tuple $BPSM = (W, R, D, P, T, I)$, where:
\begin{itemize}
    \item $W = (start, sink, N, F, t)$ is a \textbf{WF-graph};
    \item $R$ is a set of \textbf{resources} involved in executing the process;\footnote{
        Each $r \in R$ has attributes such as availability and performance. 
        A detailed formalization of these attributes is out of the scope of this paper, see e.g.~\cite{LopezPintadoDB24}.
    }
    \item $D: A(W) \rightarrow \in \mathcal{P}((0,\infty))$ is a function that maps each activity $a \in A(W)$ to a probability distribution over positive real numbers, representing the distribution of its \textbf{duration} (processing time, excluding waiting time);
    \item $P: CF(W) \rightarrow [0,1]$ is a function that assigns a \textbf{probability} to each conditional flow of $W$;
    \item $T: E(W) \rightarrow \mathcal{P}((0,\infty))$ is a function that maps each event $e \in E(W)$ to a probability distribution over positive real numbers, representing the distribution of the \textbf{waiting time} engendered by this event; and
    \item $I = \mathcal{P}((0,\infty))$ is a probability distribution over positive real numbers, representing the distribution of the \textbf{inter-arrival times} between each two consecutive case arrivals in the process.
\end{itemize}
\end{definition}

%!TEX root = main.tex
\section{Approach\label{sec:approach}}

This section begins by formalizing the state of a business process simulation, describing each of the components necessary to resume a simulation at a given point in time (\sectionname~\ref{subsec:state-formalization}).
Next, it proposes an approach to derive the state of an ongoing process given an event log of ongoing cases and a WF-graph (\sectionname~\ref{subsec:discover-state}). 
Finally, it describes the required procedure to load a state (as defined in \sectionname~\ref{subsec:state-formalization}) in a typical discrete-event simulation engine in order to resume a simulation from the given state (see \sectionname~\ref{subsec:state-loading}).

\subsection{Business Process Simulation State\label{subsec:state-formalization}}

The (discrete-event) simulation of a business process consists of the execution, by a simulation engine, of a set of events.
These events (e.g., the arrival of a new case or the completion of an activity instance) update the state of the simulated process, producing an event log with information related to such executions.
A BPS engine is typically comprised of various components that handle different aspects of a simulation (e.g., resource allocation).
With the execution of an event, each of these components updates its state.
For example, after the completion of an activity instance, the allocated resource becomes available, and the control-flow propagates through the process model, potentially enabling other activities.
We observe that BPS engines~\cite{DBLP:conf/edoc/Lopez-PintadoHD22,DBLP:journals/is/MeneghelloFGR25,DBLP:conf/bpm/PufahlW16a,DBLP:journals/dke/RozinatWAHF09} keep track of \textit{i)} the control-flow marking of each ongoing process case; \textit{ii)} the queue of activities (or events) being processed, storing their enablement and start timestamps; \textit{iii)} the queue of enabled nodes, storing their enablement timestamp; \textit{iv)} the queues of busy and available resources, storing the timestamp when they became busy and available, respectively; and \textit{v)} the state of data attributes associated to the process.
We define the state of a BPS as follows:

%We observed that a simulation engine must store the process state of each ongoing case at any simulation point in time.
%Specifically, it needs to keep track of the control-flow marking (i.e., which flows or tasks hold tokens), the timestamps at which these elements become enabled or start executing, and the resources currently occupied.
%This ensures that, when an event is processed, the simulation can correctly update the availability of resources, the enabling of tasks and events, and any branching decisions (e.g., at gateways). 
%Accordingly, the state of a simulation of a business process must store the information needed for each of these components to continue processing future events.
%To cover all these components, we propose to define the state of a BPS as follows:

\begin{defn}[State of a Business Process Simulation]\label{def:process-state}
    Let $\Phi$ be the set of case identifiers representing ongoing process instances, $W = (start, sink, N, F, t, g)$ be the workflow graph of the process, $R$ be a set of resource identifiers, and $T$ be a time domain.
    We define the \textit{state of the business process simulation at time $\tau_{st} \in T$} as the set
    %\[
    %    (\tau^{last}_{arr},\,\Xi),
    %\]
    %where $\tau^{last}_{arr}\in T$ is the timestamp of the most recent case arrival
    %recorded \emph{before} $\tau_{st}$ and  
    \[
        \Xi_{\tau_{st}}=\{\,(\phi,\,arr_{\phi},\,F_\phi,\,A_\phi,\,en_\phi,\,st_\phi,\,res_\phi)\mid \phi\in\Phi\,\}
    \]
    where, for each tuple:
    \begin{itemize}
        \item $\phi \in \Phi$ denotes the \textbf{case identifier} of an ongoing case;
        \item $arr_{\phi} \in T$ denotes the time instant when that case \textbf{arrived} at the system;
        \item $F_\phi \subseteq F$ denotes the set of \textbf{flows} in case $\phi$ that ``contain a token'';
        \item $A_\phi \subseteq N$ denotes the set of \textbf{activities} in case $\phi$ that are being processed;
        \item $en_\phi : (F_\phi \cup A_\phi) \to T$ is the \textbf{enablement function}, returning the timestamp at which a flow or activity became enabled;
        \item $st_\phi : A_\phi \to T$ is the \textbf{start function}, returning the timestamp at which an ongoing activity started being processed; and
        \item $res_\phi : A_\phi \to \mathcal{P}(R)$ is the \textbf{resource function}, returning the resource(s) allocated to an ongoing activity.
    \end{itemize}
\end{defn}

As we explain in \sectionname~\ref{subsec:state-loading}, this information is sufficient for a BPS engine to initialize all the above-mentioned queues with the information reflecting the ongoing state of the process.\footnote{Initialization of data attributes is out of scope of this paper (see \sectionname~\ref{sec:conclusion}).}

\subsection{Discovery of an Ongoing BPS State\label{subsec:discover-state}}

In this section, we present an approach to, given an event log containing ongoing cases and a WF-graph of the process, derive the state of the process at time $\tau_{st}$ ($\Xi_{\tau_{st}}$).

\begin{figure}[t]
    \vspace*{-3mm}
    \centering
    \includegraphics[width=\columnwidth]{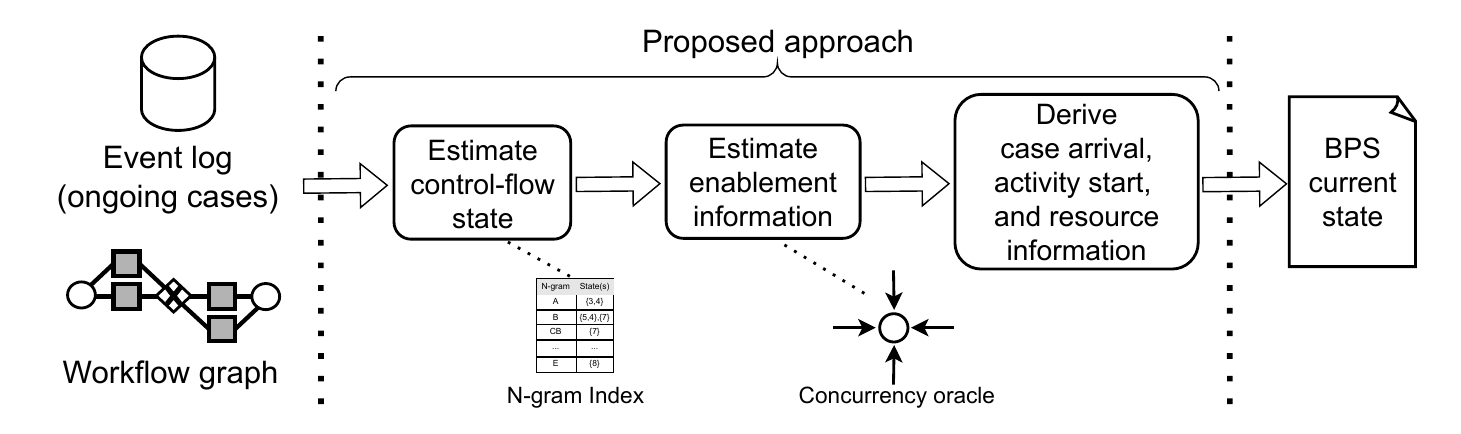}
    \caption{Overview of the proposed approach to compute the BPS current state.}
    \label{fig:approach-overview}
    \vspace*{-3mm}
\end{figure}

\figurename~\ref{fig:approach-overview} presents a high-level overview of our proposal to discover the state of an ongoing process.
Given an event log $\mathcal{L}$ of ongoing cases (see in \sectionname~\ref{sec:background}) and a WF-graph $W = (start, sink, N, F, t)$ (see \definitionname~\ref{def:workflow-graph}), our proposal is to first compute the control-flow state (i.e., marking) of each ongoing case.
Then, given the set of flows, activities, and events present in the ongoing marking, we compute the timestamp at which each of them became available.
Finally, we derive the start timestamp and resource associated with each of the ongoing activities based on the event log information, and estimate the arrival timestamp of this case.

%\smallskip
\noindent\textbf{Control-flow state.}
Computing the control-flow state (i.e., marking) of an ongoing case in a sequential process is trivial, as this is defined by the last executed activity.
However, in the case of processes with parallelism and loops, this computation becomes challenging as the last event might not be enough to denote the current marking of the case~\cite{DBLP:journals/tsc/chapela2025efficient}.
For example, the last recorded activity instance of the ongoing case 102 in \tablename~\ref{tab:event_log_interleaved} indicates that the activity ``Send Invoice'' (see \figurename~\ref{fig.bpmn-model}) is being executed, thus holding a token.
However, it gives no information whether ``Prepare package'' is enabled, ongoing, or finished.
To answer this, one must inspect previous recorded activity instances until finding either ``Prepare package'' (meaning ``Prepare package'' is ongoing or finished) or an activity prior to the AND-split that led to ``Send Invoice'' (meaning ``Prepare package'' is enabled).
Furthermore, the recorded behavior of an ongoing case may deviate from the structure defined by the input WF-graph.
This may happen not only in the case of conformance deviations, but also in cases in which the input WF-net corresponds to a ``to-be'' configuration of the process (to perform a what-if analysis).
%an analyst makes adjustments to a BPS model of an ``as-is'' process to capture a ``what-if'' scenario and then triggers a simulation of the what-if scenario starting of a log of ongoing cases. 
%In this scenario, some ongoing cases might follow pathways that deviate from the what-if model, as they do not contain the newly added task.

To estimate the current marking of an ongoing case given its trace prefix, our proposal is to adapt the technique presented in~\cite{DBLP:journals/tsc/chapela2025efficient} to the setting addressed in this paper.
This approach proposes to build an index that associates each $m$-gram (i.e., sequence of $m$ activities), representing the last $m$ executed activities of an ongoing case, with the set of markings in the WF-net that its execution leads to.
In this way, the marking of an ongoing case can be efficiently computed by querying the index with its last $m$ executed activities.

The proposal presented in~\cite{DBLP:journals/tsc/chapela2025efficient} is designed for process models represented in the Petri net formalism, and for ongoing cases logs storing one single timestamp per activity instance, thus estimating a marking either prior to or posterior to the execution of an activity.
To adapt this algorithm to WF-graphs, we propose to follow the execution semantics described in \sectionname~\ref{sec:background}, and consider an XOR-split gateway as a decision point of the process (analogously to a place with multiple outgoing silent transitions in a Petri net, see~\cite{DBLP:journals/tsc/chapela2025efficient}).
Regarding the support for non-atomic activities, our proposal is to transform the WF-graph by replacing each activity $n_i \in N \mid t(n) = task$ with a two-activity sequence, one denoting its start ($n_{i,s}$) and another one denoting its end ($n_{i,e}$).
%to WF-graphs (rather than Petri nets) and extended it to consider ongoing activities in this process.
%The proposal is designed for atomic activities, estimating a state either prior or posterior to the execution of an activity.
%To consider ongoing activities, we adapted the generation of the reachability graph proposed in~\cite{DBLP:journals/tsc/chapela2025efficient} to build an automaton where each activity generates two transitions, one associated with its start and another one associated with its end.
%In this way, after building the index, we query this map for the sequence of the last recorded events.

Following the example of case 102, the $m$-gram $\langle$``Prepare Package''$_c$, ``Prepare Invoice''$_s$, ``Prepare Invoice''$_c$, ``Send Invoice''$_s$$\rangle$, retrieving the marking in the model containing a token in the activity ``Send Invoice'' and another one in the outgoing edge of ``Prepare Package''.
This produces, for each ongoing case $\phi$, the sets $F_\phi$ and $A_\phi$, containing the sets of flows and activities with a token.

%\begin{figure*}[t]
%    \vspace*{-3mm}
%    \centering
%    \includegraphics[width=0.8\textwidth]{figures/approach.pdf}
%    \caption{Overview of the proposed approach}
%    \label{fig:approach-overview}
%    \vspace*{-3mm}
%\end{figure*}

For generating the reachability graph (see~\cite{DBLP:journals/tsc/chapela2025efficient} for details) that leads to the index, we followed a (lazy-traversal) policy that generates states prior to decision points (i.e., XOR-split gateways) when possible.
For example, the marking associated with the sequence $\langle$``Collect Customer Info.''$_s$, ``Collect Customer Info.''$_c$$\rangle$ corresponds to the outgoing flow of ``Collect Customer Information''.
In this way, the traversal of decision points is delegated to the BPS engine, which will perform such an operation based on information like the probabilities associated with each outgoing conditional flow.

While this method estimates a best-fit marking given a trace prefix, in the presence of deviations, it may incorrectly estimate that an activity is ongoing when the event log did not record its start, or vice versa.
Therefore, we post-process the marking to repair these discrepancies by:
\textit{i)} removing an activity $n_i \in A_\phi$ and adding its incoming flow in the WF-graph to $F_\phi$ if the activity is not recorded as ongoing in the input event log; or
\textit{ii)} removing the information of a recorded ongoing activity if its label $n_i \notin A_\phi$.

%To determine which flows and tasks hold tokens, we adapt the marking-reconstruction technique from \cite{DBLP:journals/tsc/chapela2025efficient}, originally intended for Petri nets. We first construct a reachability graph that captures every possible BPMN state (marking), splitting each BPMN task into a “start” node and a “completion” node so we can represent partially executed tasks. Next, for each ongoing case, we sort its logged activity instances by \texttt{StartTime} and feed this sequence to an \emph{n-gram index} \cite{DBLP:journals/tsc/chapela2025efficient}, which matches the observed tasks to states in the reachability graph. If the log shows an \texttt{EndTime}, we mark that activity as completed; if \texttt{EndTime} is \texttt{NULL}, we keep a token in that activity to mark it as ongoing. This step results in:
%\[
%F_\phi \quad(\text{flows with tokens (i.e., the ones leading to enabled elements)}),\quad
%A_\phi \quad(\text{ongoing or enabled tasks (the activities with tokens)}).
%\]

%Once $F_\phi$ and $A_\phi$ are known, we determine the time each element became enabled ($en_\phi$), 
%the start time of each ongoing activity ($st_\phi$), and the resources assigned ($res_\phi$):

%\smallskip
\noindent\textbf{Enablement information.}
%In the second stage of our proposal, we compute the enablement time ($en_\phi$) of each $f_i \in F_\phi$ and each $a_i \in A_\phi$.
In a sequential process, an activity is considered to be enabled when its chronologically preceding activity instance completes.
However, due to the existence of parallelism, the concept of enablement is redefined as the completion time ($\tau_c$) of the most recent \emph{causal predecessor} in the same case~\cite{DBLP:journals/is/ChapelaCampaD24}.
In this setting, the causal predecessor of an activity instance $\varepsilon_i$ is the most recent activity instance $\varepsilon_j$ not overlapping and such that $\alpha(\varepsilon_i)$ is not concurrent with $\alpha(\varepsilon_j)$.
To compute the concurrency relations between the activities of the process, we use the concurrency oracle proposed in~\cite{DBLP:journals/is/ChapelaCampaD24}, which considers two activities concurrent if they frequently (w.r.t.\ a given threshold) overlap in time within the same case (i.e., no strict ordering is required).\footnote{
    We consider the ongoing cases log as the source of truth to compute the enablement of the elements in the ongoing marking, as it corresponds to the ``as-is'' version of the process.
    We do not extract concurrency relations from the provided BPMN model as it constitutes the ``to-be'' version of the process and may contain concurrency relations that do not apply to ongoing cases.
}
%\footnote{
%  We note that other notions of concurrency have been proposed in the field of process mining.
%  An in-depth treatment of concurrency notions in process mining is provided by Armas-Cervantes et al. in~\cite{DBLP:journals/toit/Armas-Cervantes19}.
%  In this paper, we adopt the notion of overlapping concurrency between pairs of activities as defined above. The proposed approach is, however, modular and could be adapted to exploit other notions of concurrency.
%}
This process populates the function $en_\phi$ of the ongoing state.

%Otherwise, if an activity $A$ always finishes before $B$ starts in the same case, we treat $A$ as a causal predecessor of $B$.
%Formally, for an activity instance $\varepsilon_t$ in the log, the set of \emph{causally preceding} activities is any $\varepsilon_s$ from the same case whose \texttt{EndTime} occurs before $\varepsilon_t$’s \texttt{StartTime}, provided $\varepsilon_s$’s activity is \emph{not concurrent} with $\varepsilon_t$’s. 
%Given this relation, the earliest enablement time of $\varepsilon_t$ is generally the \texttt{EndTime} of $\varepsilon_{cp}(\varepsilon_t)$. If an activity has multiple upstream tasks (due to AND-join gateways), we take the \emph{maximum} of all those tasks’ \texttt{EndTime} values.

%\smallskip
\noindent\textbf{Arrival, start, and resource information.}
The final stage of our proposal consists of estimating the case arrival ($arr_{\phi}$), and deriving both the start timestamp ($st_\phi$) and assigned resources ($res_\phi$) for each ongoing activity.
We propose to estimate the arrival of a case $\Phi$ as its earliest start timestamp $arr_\phi = \min_{\varepsilon_i\in\Upsilon_{\phi}}\bigl(\tau_{s}(\varepsilon_i)\bigr)$.
Regarding the ongoing activities, we consider the ongoing cases log as a source of truth and retrieve, for each activity $n_i \in A_\phi$, the recorded start timestamp ($st_\phi(\varepsilon_i)$) and recorded resource(s) ($res_\phi(\varepsilon_i)$) of the latest event in the case recording the execution of $\alpha$.

\subsection{Short-term Simulation from Current State\label{subsec:state-loading}}

In this section, we describe the required procedure to load a given state $\Xi_{\tau_{st}}$ (as defined in \sectionname~\ref{subsec:state-formalization}) in a typical discrete-event simulation engine in order to resume a simulation of a BPS model $BPSM = (W, R, D, P, T, I)$ from such a state.
We present this process over the BPS engine Prosimos~\cite{DBLP:conf/edoc/Lopez-PintadoHD22}.
However, we foresee that this process might be generalized to other (discrete-event) BPS engines such as the ones presented in~\cite{DBLP:journals/is/MeneghelloFGR25,DBLP:conf/bpm/PufahlW16a,DBLP:journals/dke/RozinatWAHF09}.
%The input needed for a BPS engine to perform a short-term simulation starting at a time $\tau_{st}$ is composed of the BPS model $BPSM = (W, R, D, P, T, I)$ (as defined in \sectionname~\ref{def:bps-model}) and the process ongoing state $(\\tau^{last}_{arr},\\,\\Xi_{\\tau_{st}})$ (as defined in \definitionname~\ref{def:process-state}).

%\smallskip
\noindent\textbf{Control-flow initialization.}
%The first element to be initialized in the BPS engine is the control-flow state.
The first step is to load the marking of each case $\phi$ composed of the flows in $f \in F_\phi$ and the activities $a_i \in A_\phi$.
As mentioned in \sectionname~\ref{subsec:discover-state}, the flows composing $F_\phi$ may be enabling gateways, events, or activities.
For this reason, prior to continuing with the other components, the engine must process the advancement of these flows according to the following reasoning: 
\textit{(1)} If an activity $a_i \in A(W)$ is enabled by a flow $f_i \in F_\phi$, the engine adds $a_i$ to the set of enabled activities, setting its enabled time to $en(f_i)$.
\textit{(2)} If an event $e_i \in E(W)$ is enabled by a flow $f_i \in F_\phi$, the engine simulates the duration of the event with $T(e_i)$ and subtracts from it the period that has already happened since the flow was enabled ($dur_{e_i} = T(e_i) - (\tau_{st} - en(f_i))$); if $dur_{e_i} >= 0$, the completion of $e_i$ is queued with end time $\tau_{st} + dur_{e_i}$ for future processing; otherwise, if $dur_{e_i} < 0$, $e_i$ is assumed to have finished at $\tau_{st} + dur_{e_i}$, and the engine keeps advancing through the outgoing flow of $e_i$ updating its enablement time to be the completion of $e_i$.
\textit{(3)} If an XOR-split gateway $g$ is enabled by a flow $f_i \in F_\phi$, the engine traverses $g$ according to the probabilities of its outgoing flows assigned by $P$.
This traversal step is repeated until all enabled elements are either already queued activities, or in-process events.

%\smallskip
\noindent\textbf{Ongoing activity initialization.}
%The next step is to load the ongoing activities denoted in $A_\phi$.
For each $a_i \in A_\phi \mid \phi \in \Phi$, the BPS engine computes its duration with $D(a_i)$, subtracts from it the period that has already happened since the activity started ($dur_{a_i} = D(a_i) - (\tau_{st} - st(a_i))$),\footnote{
  If the resource performing an ongoing activity has an associated availability calendar, we compute the duration of the activity relative to this calendar, i.e., discarding the off-duty period from the duration computations.
}
and queues its completion at $\tau_{st} + \max(0,\;dur_{a_i})$ for future processing.
The duration of ongoing activities with a recorded resource not present in the BPS model (e.g., due to a change in the what-if configuration) is computed considering the first resource capable of performing that task.
For the ongoing activities with no recorded resource, the BPS model assigns a new resource from the queue of available resources capable of performing that task, and computes its duration accordingly.\footnote{
    The resource allocation strategy proposed in this paper could be extended to consider more advanced criteria, as discussed in~\cite{arias2017towards}.
}

%\smallskip
\noindent\textbf{Resource initialization.}
For each $r_i \in res_\phi(a_i) \mid a_i \in A_\phi \wedge \phi \in \Phi \wedge r_i \in R$ (i.e., the resources recorded to be performing an ongoing activity, such that they are part of the resources of $BPSM$), the BPS engine adds the resource $r_i$ to the queue of busy resources with $st_\phi(a_i)$ as the timestamp they stopped being available for processing.

%\smallskip
\noindent\textbf{Case arrival initialization.}
To queue the arrival of the next case, the BPS engine retains the latest case arrival $\tau_{arr}$ in $\Xi_{\tau_{st}}$, computes an inter-arrival duration $ia$ with $I$, and queues the first future case at $\max(\tau_{st}, \tau_{arr} + ia)$.
Subsequent arrivals are scheduled recursively when processing each new arrival.\footnote{
    Alternative case arrival model implementations (e.g., defining a Poisson distribution for different week/day time intervals) may use other $\tau_{arr}$.
}

%\smallskip
\noindent\textbf{Short-term simulation.}
After loading all the components, the BPS engine resumes the simulation by processing the queued future completion events, as well as starting the execution of the queued enabled activities, when possible.
Due to the availability of enablement information, this process can be performed following the priority policies (e.g., FIFO or LIFO) implemented in the simulation engine.

\begin{figure}[t]
    \vspace*{-3mm}    
    \centering    
    \includegraphics[width=0.5\textwidth]{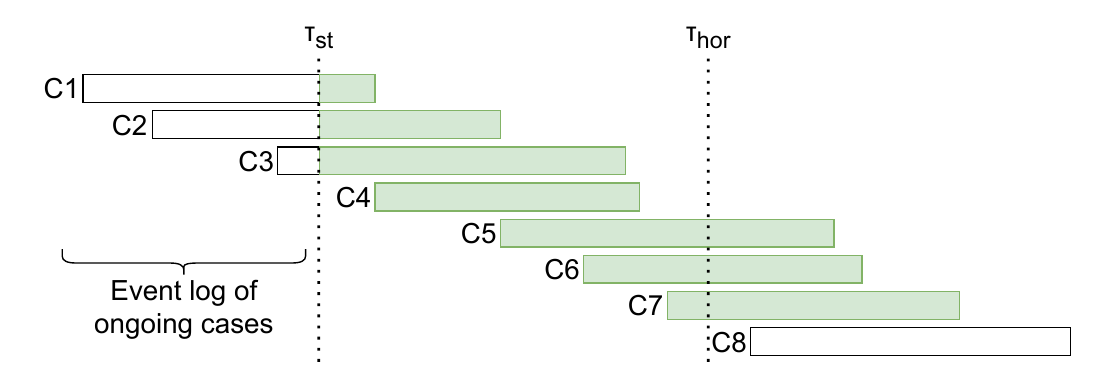}
    \caption{Short-term simulation outcome.}
    \label{fig:short-term-output}
    \vspace*{-3mm}    
\end{figure}

Our proposal for a short-term simulation consists of, given a simulation horizon $\tau_{hor}$ defining the end of the simulation, simulating the execution of the process until all cases that arrived prior to $\tau_{hor}$ are completed.
Cases that begin strictly after $\tau_{hor}$ might still be simulated to ensure realistic resource contention.
Then, the simulated event log is the result of retaining the activity instances of cases that arrived prior to $\tau_{hor}$.
Figure~\ref{fig:short-term-output} offers a simplified illustration of this process.

% In a \emph{short-term} scenario, the engine runs up to a horizon $\tau_{st} + d$ and collects results for cases that 
% \begin{enumerate}
%     \item Either started before or at $\tau_{st}$ (the process state), or
%     \item Arrive in $[\tau_{st}, \tau_{st}+d]$.
% \end{enumerate}
% 
% Figure~\ref{fig:short-term-output} offers a simplified illustration. Cases C1, C2, and C3 began before $\tau_{st}$ and continue as part of the process state. 
% Case C4 starts after $\tau_{st}$ yet completes within the horizon. 
% Cases C5, C6, and C7 also start after $\tau_{st}$, but finish beyond the horizon; they remain in simulation to reflect ongoing resource usage, 
% while their final events may be discarded from the final log. 
% Case C8 arrives entirely after the horizon, so although it can be simulated for resource contention, 
% it contributes nothing to the final output. 

% By initializing the simulation engine with the \emph{real-time} process state, we ensure that ongoing tasks, incomplete gateways, and resource allocations match the actual conditions at $\tau_{st}$. 

%!TEX root = main.tex

\section{Evaluation\label{sec:evaluation}}

This section presents an experimental evaluation to analyze the accuracy of initializing a short-term simulation from the ongoing state, as proposed in this paper, in comparison with a warmed-up state, using both synthetic and real-life processes.

To this end, the first part of the evaluation addresses the following question:
\textit{Evaluation Question 1 (EQ1): Does starting the short-term simulation from the ongoing state improve its accuracy in event logs where the workload is unstable?}
To answer this, we designed a set of simulation scenarios with varying levels of complexity and workload patterns. % (see \sectionname~\ref{sec:synthetic-eval}).
The second part of the evaluation focuses on the following question:
\textit{Evaluation Question 2 (EQ2): How does our approach perform with real-life processes, where the BPS model may be less accurate?}
To investigate this, we analyzed three event logs from real-life processes: one with a stable workload and two exhibiting workload concept drift. % (see \sectionname~\ref{sec:real-life-eval}).

\subsection{Synthetic evaluation\label{sec:synthetic-eval}}

This section presents the evaluation to assess if starting the short-term simulation from the ongoing state improves its accuracy in event logs where the workload is unstable (EQ1).

%\smallskip
\noindent\textbf{Datasets and Setup.}
To evaluate EQ1, we selected the BPS models of two synthetic processes of different complexities: 
one of a Purchase-to-Pay process composed of 24 activities, four loops, and three skippable activities (simple structure);
and another one of a Loan Application process composed of 17 activities, exhibiting two parallel structures (with three and two branches), three loops, and five decision points (complex structure).
For each of these processes, we created three configurations: 
one of them with a \emph{stable} workload throughout the process, by assigning a constant arrival rate and 24/7 availability calendars;
another one with a workload that fluctuates over time following a \emph{circadian} pattern on a weekly basis, by changing the availability calendars to Mon-Fri 9-17;
and a last one with an \emph{unstable} workload, by combining the simulation of different configurations of the process, creating multiple concept-drift points within the log.

To avoid a bias in the results due to using the same simulation engine, we generated two logs for each of these process configurations with APROMORE.\footnote{\url{https://apromore.com/}}
Then, we discovered a BPS model with one of these logs by using SIMOD~\cite{DBLP:journals/softx/ChapelaCampaLSD25} (v5.1.5), and used the other log as the ground truth log.

For each ground truth log, we selected three points in the log timeframe in which the workload (number of ongoing cases) was similar to 10\%, 50\%, and 90\% of the maximum workload of the entire log, representing the start of each short-term simulation ($\tau_{st}$).
We retained the cases ongoing at each $\tau_{st}$ (as the ongoing cases log), and ran a short-term simulation with a horizon equal to the 90$^{th}$ percentile of the case durations of that log.
By selecting the 90$^{th}$ percentile of the case durations, we discard long outlier durations while ensuring that most cases that were ongoing at $\tau_{st}$, as well as some that start shortly after, will be completed by the end of the simulation.\footnote{A simulation with a longer horizon, effectively including a higher proportion of complete cases, must be considered as a long-term simulation.}

As a baseline, we use a warm-up approach ($WarmUp$), which, starting from an empty state, creates and simulates cases according to the simulation model given as input, until the number of ongoing cases in the simulation is equal to the actual number of ongoing cases as of time $\tau_{st}$ (meaning that the workload in the simulation model reflects the current state), or until the simulation has lasted for a maximum warm-up period of $\tau_{hor} - \tau_{st}$.
%We do not compare our proposal with the techniques presented in~\cite{DBLP:conf/bpm/WynnDFHA07,DBLP:journals/dke/RozinatWAHF09} because these techniques require %the enablement time for each activity instance in each ongoing case. This information can be extracted if the process is executed in a workflow management system, or a similar system that detects and records activity enablement events.
%a workflow management system, or a similar system that detects and records activity enablement events.
%However, this information is not available when the log comes from a system of record, such as an ERP or CRM system, which are typical sources of event logs for process mining.
%We also do not use the technique in \cite{DBLP:conf/bpm/PourbafraniLLA23} as a baseline because it does not calculate the enablement time of activity instances of ongoing cases, nor does it detect which activity instances are ongoing and which are not.
%We also do not use the technique in \cite{DBLP:conf/bpm/PourbafraniLLA23} as a baseline because it does not calculate which activity instances are enabled for an ongoing case, nor their enablement time.
%This is necessary in order to initialize the queues of activity instances that will be allocated to the resources in the process (see \sectionname~\ref{sec:related-work}).

We measure the goodness of a short-term simulation (averaged across 10 simulation runs) by means of three distance metrics, which capture discrepancies between the output of the short-term simulation and the ground-truth log of ongoing cases:
\textit{(1)} The absolute difference in the number of ongoing cases (OCD) at $\tau_{st}$; this metric captures how well the simulation reflects the workload of the process.
\textit{(2)} The N-gram Distance (NGD)~\cite{DBLP:journals/is/ChapelaCampaBBDKS25} between the simulated cases (as described in \figurename~\ref{fig:short-term-output}) and the analogous cases in the ground truth log; this metric reflects how well the simulation mirrors the actual sequences of activities observed in the ground truth.
%Before applying NGD, we preprocess the event logs to retain only activity instances that started or finished within the simulation window.\footnote{
%  Note that this filtering produces incomplete cases. However, as the measure is not based on the complete sequence of activity instances, but on each pair of activity instances, its validity still holds.
%} 
\textit{(3)} The Remaining Cycle Time Distance (R-CTD), a variation of the Cycle Time Distance measure proposed in~\cite{DBLP:journals/is/ChapelaCampaBBDKS25} that compares the distribution of the remaining times of the cases ongoing as of time $\tau_{st}$. Specifically, the R-CTD distance between a simulated log and an actual log is the earth-movers distance between the distribution of remaining times of ongoing cases produced by the simulation and the same distribution of remaining times extracted from the ground-truth log.

\begin{table}[t]
    \centering
    \scriptsize
    \caption{
      Results of the synthetic evaluation.
    }
    \label{tab:synthetic-evaluation}

    \vspace*{-3mm}
    \setlength\tabcolsep{7pt}
    \begin{tabular}{l l r r r}
        \toprule
                &            & \multicolumn{1}{c}{OCD} & \multicolumn{1}{c}{NGD} & \multicolumn{1}{c}{R-CTD} \\ \toprule \toprule

        P2P$_{stable}$        & $WarmUp$     & 2.50            & \colorcell0.06 & \colorcell2.18  \\
                              & $ProcState$  & \colorcell0.00  & \colorcell0.06 & 3.09            \\ \midrule

        P2P$_{circadian}$     & $WarmUp$     & 3.30            & \colorcell0.08 & \colorcell5.78  \\
                              & $ProcState$  & \colorcell0.00  & \colorcell0.08 & 7.58            \\ \midrule

        P2P$_{unstable}$      & $WarmUp$     & 18.30           & 0.17           & 29.53           \\
                              & $ProcState$  & \colorcell0.00  & \colorcell0.14 & \colorcell15.63 \\ \midrule
                              
        LoanApp$_{stable}$    & $WarmUp$     & 6.90            & 0.30           & 3.30            \\
                              & $ProcState$  & \colorcell0.00  & \colorcell0.24 & \colorcell3.25  \\ \midrule

        LoanApp$_{circadian}$ & $WarmUp$     & 9.67            & 0.45           & 29.33           \\
                              & $ProcState$  & \colorcell0.00  & \colorcell0.31 & \colorcell10.21 \\ \midrule

        LoanApp$_{unstable}$  & $WarmUp$     & 1.00  & 0.63           & 47.50           \\
                              & $ProcState$  & \colorcell0.00  & \colorcell0.60 & \colorcell31.32 \\ \midrule
    \end{tabular}
    \vspace*{-3mm}
\end{table}

%\smallskip
\noindent\textbf{Results and Discussion.}
Regarding EQ1, \tablename~\ref{tab:synthetic-evaluation} shows the results of the synthetic evaluation, as the average of the three above-mentioned starting points.
$ProcState$ starts the simulation with a number of ongoing cases (OCD) similar to the real one, while $WarmUp$  fails to consistently reach this level in certain scenarios.
The results regarding the NGD show that starting the simulation at the ongoing state ($ProcState$) leads to a better replication of the control-flow.
Finally, with respect to the ability to replicate the temporal perspective (R-CTD), as expected, $WarmUp$ and $ProcState$ have comparable results in logs with a stable workload.
The difference grows when the workload is stable but follows a circadian trend, where they still perform similarly in the structurally simple process ($P2P$).
Finally, $ProcState$ presents better results in the circadian configuration of the structurally complex process ($LoanApp$), and in both cases when the workload is unstable.

%However, in terms of temporal metrics (\emph{AED} and \emph{R-CTD}), our \textit{ProcState} method outperforms the baselines more clearly. Notably, in the \emph{stable} arrival scenarios, the gap in R-CTD grows.
%A possible explanation is that the warm-up approaches have difficulty replicating the true remaining durations in a system that has already been running steadily, whereas \textit{ProcState} inherits the actual progress of each ongoing case.
%This leads to more accurate completion time predictions for those cases, improving the overall distribution of remaining cycle times. Moreover, in \emph{wobbly} scenarios with more unpredictable start times, \textit{ProcState} achieves a closer match in AED.
%This may come from \textit{ProcState} inheriting the true in-progress tasks and waiting periods directly from the log, while the warm-up baselines only approximate them.

% {\color{blue}[\textit{Here is basically adding a table with the results, and commenting about them. You can add reasonings and explanations on why the results are like that along with them, or at the end of this paragraphs altogether.}]}

% {\color{blue}[\textit{Again, take a read at the ``result'' part of the N-gram index paper, and try to follow a similar structure, in the end is same abstract structure.}]}

%\todo[inline]{data as requested}
%cases in the ALog:
%loan stable = 46
%loan wobbly = 10
%p2p stable = 45
%p2p wobbly = 53

\subsection{Real-life evaluation\label{sec:real-life-eval}}

This section presents the evaluation to assess whether starting the short-term simulation from the ongoing state improves its accuracy in the case of real-life processes (EQ2).

%\smallskip
\noindent\textbf{Datasets and Setup.}
To assess EQ2, we selected three real-life datasets of different complexities:
%a log from an academic credentials management process (AC\_CRE), containing a high number of resources exhibiting low participation in the process;
two logs of loan application processes from the Business Process Intelligence Challenges (BPIC) of 2012\footnote{\url{https://doi.org/10.4121/uuid:3926db30-f712-4394-aebc-75976070e91f}} and 2017,\footnote{\url{https://doi.org/10.4121/uuid:5f3067df-f10b-45da-b98b-86ae4c7a310b}} preprocessed as described in~\cite{DBLP:journals/is/ChapelaCampaBBDKS25};
and a log of a purchase-to-pay process (WORK\_O).\footnote{
We excluded the call center and academic credentials management event logs used in~\cite{DBLP:journals/is/ChapelaCampaBBDKS25}.
The call center log has less than two activity instances per case on average, and thus, it is not suitable to evaluate the quality of a technique that computes the current state of a case.
The academic credentials log contains less than 1000 long-duration cases. After partitioning the data, there are not enough samples to train a BPS model and then test the quality of a short-term simulation against the second part of the log.}
We preprocessed each of these logs by adding artificial start and end events to each case.
\tablename~\ref{tab:log-characteristics} shows the characteristics of the three event logs.
While BPIC12 shows more stable workloads throughout the entire log, WORK\_O and BPIC17 present a concept drift with a change in the demand in the middle of the process execution.
We split these datasets into two event logs of complete cases (i.e., training and test partitions) as recommended in~\cite{DBLP:journals/is/ChapelaCampaBBDKS25}.
In the case of WORK\_O and BPIC17, the split was performed at the concept-drift point.
We discovered the optimized BPS model with SIMOD~\cite{DBLP:journals/softx/ChapelaCampaLSD25} (v5.1.5) from the training partition.
The procedure to define the simulation horizon, the considered baseline, and the used measures of goodness is similar to that used to assess EQ1.
Regarding the timestamps when to start the simulation ($\tau_{st}$), we divided each log into 10 intervals of equal size (discarding the first and last periods equal to the simulation horizon) and randomly selected a timestamp within each of these intervals ($\tau_{st}$).
We then run 10 short-term simulations for each $\tau_{st}$.

\begin{table}[t]
    \centering \scriptsize
    \caption{Characteristics of the real-life logs used in the evaluation.}
    \label{tab:log-characteristics}

    \vspace*{-3mm}
    \setlength{\tabcolsep}{5pt}
    \begin{tabular}{l r r r r r}
        \toprule
                     \multicolumn{1}{c}{Event log} & \multicolumn{1}{c}{Cases} & \multicolumn{1}{c}{Activity instances} & \multicolumn{1}{c}{Variants} & \multicolumn{1}{c}{Activities} & \multicolumn{1}{c}{Resources} \\ \toprule \toprule
        %AC\_CRE  & 954   & 4,962 & 97 & 16 & 559 \\ \midrule
        BPIC12   & 8,616 & 59,302 & 2,115 & 6 & 58 \\ \midrule
        WORK\_O  & 19,798 & 149,664 & 1,026 & 24 & 201 \\ \midrule
        BPIC17   & 30,276 & 240,854 & 6,358 & 8 & 148 \\ \bottomrule
    \end{tabular}
    \vspace*{-3mm}
\end{table}

\begin{comment}
\begin{table}[t]
    \centering
    \scriptsize
    \caption{
      Results AVERAGE for the real-life datasets. OCCD - ongoing cases count difference, RECD - remaining event per case difference, 90th percentile.
    }
    \label{tab:real-life-eval}

    \setlength\tabcolsep{5pt}
    \begin{tabular}{l l r r r r}
        \toprule
                &                & \multicolumn{1}{c}{OCCD} & \multicolumn{1}{c}{NGD} & \multicolumn{1}{c}{R-CTD} & \multicolumn{1}{c}{RECD} \\ \toprule \toprule

        AC\_CRED  & $WarmUp$     & 1.000 & 0.874 & 213.813 & 0.637 \\
                  & $ProcState$  & 0.000 & 0.970 & 243.538 & 0.738 \\ \midrule

        BPIC\_12  & $WarmUp$     & 259.390 & 0.495 & 161.377 & 0.297 \\
                  & $ProcState$  & 0.000 & 0.518 & 113.375 & 0.458 \\ \midrule

        BPIC\_17  & $WarmUp$     & 202.780 & 0.334 & 60.322 & 0.080 \\
                  & $ProcState$  & 0.000 & 0.333 & 105.057 & 0.379 \\ \midrule

        W\_ORDS   & $WarmUp$     & 1.000 & 0.427 & 121.036 & 1.043 \\
                  & $ProcState$  & 0.000 & 0.441 & 49.734 & 0.214 \\ \midrule
    \end{tabular}
    \vspace*{-3mm}
\end{table}
\end{comment}

\begin{table}[t]
    \centering
    \scriptsize
    \caption{
      Results of the real-life evaluation.
    }
    \label{tab:real-life-evaluation}

    \vspace*{-3mm}
    \setlength\tabcolsep{7pt}
    \begin{tabular}{l l r r r}
        \toprule
                &                & \multicolumn{1}{c}{OCD} & \multicolumn{1}{c}{NGD} & \multicolumn{1}{c}{R-CTD} \\ \toprule \toprule

        %AC\_CRE   & $WarmUp$     & 1.00             & \colorcell0.87 & \colorcell213.81    \\
        %          & $ProcState$  & \colorcell0.00   & 0.97           & 243.54              \\ \midrule

        BPIC12    & $WarmUp$     & 259.39           & \colorcell0.50 & 161.38              \\
                  & $ProcState$  & \colorcell0.00   & 0.52           & \colorcell113.38    \\ \midrule

        WORK\_O   & $WarmUp$     & 1.00             & \colorcell0.43 & 121.04              \\
                  & $ProcState$  & \colorcell0.00   & 0.44           & \colorcell49.73     \\ \midrule

        BPIC17    & $WarmUp$     & 202.78           & \colorcell0.33 & \colorcell60.32     \\
                  & $ProcState$  & \colorcell0.00   & \colorcell0.33 & 105.06              \\ \midrule
    \end{tabular}
    \vspace*{-3mm}
\end{table}

%\smallskip
\noindent\textbf{Results and Discussion.}
Regarding EQ2, \tablename~\ref{tab:real-life-evaluation} shows that $ProcState$ always starts the simulation with the correct number of ongoing cases, while $WarmUp$ fails to replicate the workload in BPIC12 and BPIC17.
Both approaches show comparable results w.r.t\ the control-flow metric (NGD).
Regarding the temporal metric (R-CTD), $ProcState$ presents more accurate results in BPIC12 and WORK\_O, but $WarmUp$ achieves better results in BPIC17.
By inspecting the simulated logs, we identified that the presence of very long extraneous delays (not well captured in the BPS model) caused $ProcState$ to immediately start the execution of the enabled activity instances in one single go at the start of the simulation. This is because, according to the BPS model, these activity instances should have been started already prior to $\tau_{st}$.
This shortens the remaining cycle time of the ongoing cases, giving an advantage to $WarmUp$. %, which arrives at $\tau_{st}$ with more evenly distributed cases.
This hints at a potential area for improvement for $ProcState$, which could benefit from a different policy when handling events enabled at $\tau_{st}$.

\medskip\noindent\textbf{Limitations.}
The proposed approach is designed over BPS models modeling the control-flow through BPMN models.
Although its adaptation to Petri nets might be straightforward, a different approach would be required to support other control-flow representations (e.g., $n$-order Markov chains).
Moreover, we do not incorporate the data perspective of the process due to the premature state of data-ware BPS engines.
In addition, if some resources in the event log do not appear in the BPMN model, we rely on a fallback assignment strategy, which may reduce accuracy when simulating resource usage.
We leave out of the scope of this research the data perspective of BPS models (data attributes) and plan to implement this in the future. 
Applicability of this research is also restricted due to supporting only exclusive gateways of BPS models.

%\subsection{Threats to Validity}

% {\color{red}
% \textit{[Take classic paragraph from other papers and adapt.]}
% }

\medskip\noindent\textbf{Threats to Validity.}
The reported evaluation is potentially affected by the following threats to validity.
\textit{External validity}, the experiments rely only on six simulated and three real-life processes.
The results could be different for other datasets.
%Second, regarding \textit{construct validity}, in the evaluation, we used a set of measures of goodness based on discretized distributions and time series.
%The results could be different for other measures.
\textit{Ecological validity}, the evaluation is based on the performance of as-is BPS models.
This allows us to measure how well the approach replicates the as-is process, but not to assess the goodness of the simulation models in a what-if setting, e.g., predicting the performance of the process after a change.

%We tested our approach on synthetic data and a limited set of real-life logs. Although we have designed the experiments to represent various workload conditions (steady vs.\ bursty arrivals, different concurrency levels), the results may not generalize to all real processes, especially those with extensive data dependencies or complex resource structures. 

%Our metrics (e.g., N-gram Distance, Absolute Event Distribution, Remaining Time Distribution) focus on timing and sequence aspects. These metrics do not capture all process dimensions (e.g., specific data-driven branching). Furthermore, we rely on concurrency oracles computed from logs, which could be inaccurate if the real concurrency changed significantly over time but was not reflected in the historical data.

%Since we consider the event log as ground truth for ongoing cases, any noise or missing data in the log (e.g., incomplete resource labeling) may propagate to the discovered state. Moreover, our short-term evaluation relies on partial-state initialization in a single simulation engine (Prosimos). While this setup demonstrates feasibility, different engines may handle concurrency, resource calendars, or arrival processes differently, potentially influencing outcomes.

%Despite these limitations, we believe our methodology provides a practical foundation for accurately resuming short-term simulations in BPMN-based environments.

%!TEX root = main.tex

\section{Conclusion\label{sec:conclusion}}

We presented an approach to discover the current state of a process from an event log of ongoing cases.
The approach builds upon an existing approach to compute the control-flow state of an ongoing case from its trace prefix. From there, we identify which BPMN elements (activities, events, gateways) are enabled in each case, and their enablement time. This information, combined with information about ongoing activity instances in the log, allows us to run simulations to predict the short-term performance of the process from the current state, both when the model remains unchanged (``as-is'') and when the model is altered (``what-if'').

%to support operational decisions. 
%starting from a state that reflects the current state of the resources and their queues.

%Our technique builds on a reachability graph for the given BPMN model and uses an n-gram index to match each case’s recent activity sequence against potential states in constant (or near-constant) time. After establishing which tasks are ongoing (or enabled), we derive when each task became enabled, its start time, and its assigned resources. By resuming the simulation from this process-state snapshot, we avoid the need for an artificial warm-up period or a fixed assumption of steady state. 

%Our synthetic evaluation suggests that even when processes include longer durations and parallel tasks, our process-state approach improves the simulation's ability to replicate the ground-truth distribution of timestamps and completion times.

The experimental evaluation shows that simulations started from the current state lead to more accurate forecasts of short-term performance relative to an approach where the simulation is warmed up until reaching a steady state.
The advantages of initializing the current state are particularly visible in processes with unstable performance patterns (e.g.\ high workload on some days, lower on others) and in the presence of concept drift.
Experiments with real-world event logs partially support these claims, but also suggest that the proposed state initialization leads to inaccuracies when the log contains long extraneous delays, i.e., long delays that cannot be attributed to resource contention or unavailability.
This observation highlights a limitation of the existing approach, which is that it may predict that an enabled activity instance should have been started before the current time point (i.e., before the start of the simulation).
Addressing this limitation is a direction for future work.

Another limitation of the proposal is that it does not initialize the state of data attributes. Data attributes may affect the routing of cases, and thus their remaining time and the workload they generate. A challenge here is that the data attributes in the log may differ from those allowed in the BPS model, especially when the BPS model captures a ``what-if'' scenario. Dealing with mismatches between ``as-is'' and ``what-if'' data schemas calls for schema mapping techniques.

Yet another limitation of the approach is that it is limited to workflow graphs with exclusive and parallel gateways. It does not support BPMN models with inclusive gateways or event-based gateways, among other constructs. Addressing this limitation is another future work avenue.

\medskip\noindent\textbf{Reproducibility.}
The datasets and detailed evaluation results are available at: {\small \url{https://doi.org/10.5281/zenodo.15552941}}.
The scripts to reproduce the experiments are available at: {\small \url{https://github.com/AutomatedProcessImprovement/ongoing-bps-state}}

\medskip\noindent\textbf{Acknowledgments.}
This work has been funded by the Estonian Research Council (PRG1226).

\bibliographystyle{ieeetr}
\bibliography{references}

\end{document}